\documentstyle[12pt, epsfig]{article}
\begin{document}     
\begin{flushright}
UR-1533
\end{flushright}
\begin{center}
{\bf Casimir Energy of a Spherical Shell}\\

\bigskip

M. E. Bowers and C. R. Hagen$^*$\\

\bigskip

Department of Physics and Astronomy\\
University of Rochester\\
Rochester, New York 14627\\

\end{center}

\begin{abstract}
The Casimir energy for a conducting spherical shell
of radius $a$ is computed using a
direct mode summation approach.  An essential ingredient is the implementation
of a recently proposed method based on Cauchy's theorem for an evaluation of
the eigenfrequencies of the system.  It is shown, however, that this earlier
calculation uses an improper set of modes to describe the waves
exterior to the sphere. Upon making the necessary corrections 
and taking care to ensure that no
mathematically ill-defined expressions occur, the technique is shown to leave
numerical results unaltered while avoiding a longstanding
criticism raised against earlier calculations of the Casimir energy.
\end{abstract}

\bigskip

\begin{center}
{\bf I. Introduction}\\
\end{center}

In 1948 Casimir [1] first predicted that two infinite parallel plates in 
vacuum would attract each other.  This quite remarkable result depends only on
the universal constants $\hbar$ and $c$ and the geometry of the plate
configuration, but is independent of such things as the electric
charge.  It has its origin in the expression for the energy $E$ of the
uncoupled electromagnetic field which is well known to have the form
$$E = \sum_{\bf k,\lambda}({1\over 2} + n_{\bf k,\lambda})\hbar\omega_{\bf
k}^{(\lambda)} $$

\noindent where $\omega_{\bf k}^{(\lambda)}=|{\bf k}|$ and 
$n_{\bf k,\lambda}$ denotes the  photon occupation number in the mode with wave
number ${\bf k}$ and polarization $\lambda$.  The sum is to be taken over all
allowed ${\bf k}$ and $\lambda$. In the absence of radiation this reduces to 
the zero point vacuum energy, or 

\begin{equation}
E = \sum_{\bf k,\lambda}{1\over 2}\hbar\omega_k^{(\lambda)}.
\end{equation}

\noindent This divergent expression can be rendered finite by use of a cutoff
or convergence factor, and (somewhat surprisingly) made to yield measurable
effects when boundary surfaces are used to modify the allowed set of modes
which are to be included in (1). Thus Casimir found that there is a net
attractive force between two parallel conducting plates.  This result led,
naturally enough, to his subsequent suggestion [2] that such forces could play 
a significant role in lending stability to the electron.  However, the Casimir
energy for the sphere was found to be positive, thereby implying a repulsive
force rather than the anticipated attractive one.  
This result was first achieved
in a remarkable, albeit intricate, calculation by Boyer [3] and was later
verified by a number of authors [4-6]. 

More recently a direct mode summation approach to the problem of calculating
the Casimir energy of a conducting spherical shell has been advanced by
Nesterenko and Pirozhenko [7].  Their technique uses Cauchy's theorem to
convert the sum over eigenmodes into an integral and was found to yield the
same analytic expression for the Casimir energy as that obtained in refs. 5 and
6. 
Since the original approach to the parallel plate geometry was also based on a
direct mode summation, such a result provides a welcome addition to the
literature of this subfield.  On the other hand the
approach of ref. 7 has some shortcomings which prevent it from being
immediately accepted as the natural extension of the Casimir method to the
sphere.  In particular it is shown in the following section that the standing
waves exterior to the sphere have not been found correctly in ref. 7.  After
finding appropriate expressions for those modes the calculation is put on a
more rigorous footing by including a cutoff function to dampen the high
frequency contributions (in direct analogy to the Casimir calculation [1])
together with a modification of the contour in order to ensure the convergence
of the Cauchy integral expression.  It is shown that this eliminates
a spurious divergence which Candelas [8] asserts must be added to the
result of ref. 6.  The approach developed here applies equally well to
the Dirichlet and Neumann spheres, but no explicit calculations are carried
out for these cases since earlier numerical work is unaffected.
An extension to (2+1) dimensional QED is considered in a subsequent section
with the result that there is an intrinsic divergence in the Casimir energy in
this case.

\medskip
                       
\begin{center}
{\bf II. Mode Summation for a Spherical Shell Using Cauchy's Theorem}\\      
\end{center}

In evaluating the Casimir energy of a sphere it should be noted
that each mode is $2l+1$ fold degenerate.  Thus (1) becomes

\begin {equation}
E_c=\sum_{l=1}^{\infty}(l+{1\over2})\sum_{n=1}^{\infty}\sum_{\lambda=1,2}
\omega_{nl}^{(\lambda)}
\end {equation}

\noindent where the eigenfrequencies $\omega_{nl}^{(\lambda)}$ are determined
by imposing appropriate boundary conditions on the multipole
fields.  In particular, for the case of a spherical shell of radius $a$, 
the transverse electric (i.e., $\lambda=1$) modes are

\begin {equation}
j_l(\omega a)=0
\end {equation}    

\begin {equation}
A_lj_l(\omega a)+B_ln_l(\omega a)=0
\end {equation}
and the transverse magnetic (i.e., $\lambda=2$) modes are

\begin {equation}
{d\over dr}[rj_l(\omega r)]\biggr|_{r=a}=0
\end {equation}

\begin {equation}
{d\over dr}\biggl[r[C_lj_l(\omega r) +D_ln_l(\omega r)]\biggr] \biggr| _{r=a}=0
\end {equation}
where $j_l$ and $n_l$ are the spherical Bessel functions.  The constants
$A_l$, $B_l$, $C_l$, and $D_l$ (more specifically, the ratios $B_l/A_l$,
$D_l/C_l$) are determined by prescribing the correct asymptotic behavior at
large $r$.  It should be noted that Eqs. (3) and (5) determine the interior
($r<a$) modes while (4) and (6) specify the exterior ($r>a$) modes.

The coefficients which appear in the exterior mode equations are determined
by enclosing the entire system
within a second concentric conducting sphere of radius $R$.  It is 
straightforward
to verify that Eqs. (4) and (6) can then be written as 
                         
$$j_l(\omega a)+tan\delta_ln_l(\omega a)=0$$
and
                                                     
$${d\over dr}\biggl[ r[j_l(\omega r)-cot\delta_l n_l(\omega r)]\biggr]
\biggr|_{r=a}=0$$
respectively, where for sufficiently large $R$ 

\begin{equation}
\delta _l=\omega R-{l\pi \over 2}.
\end{equation}
This is to be contrasted with the exterior mode solutions of ref. 7 where Eqs.
(4) and (6) have been replaced by those of Stratton [9], namely
$$h_l^{(1)}(\omega a)=0$$
and
$${d\over dr}[rh_l^{(1)}(\omega r)]\biggr|_{r=a}=0,$$
where $h_l^{(1)}$ denotes the Hankel function
$$h_l^{(1)}(z)=j_l(z)+in_l(z).$$
While such boundary conditions are appropriate to determine the
complex frequencies for a radiating sphere, they are not
valid to specify the necessarily real frequencies which
contribute to the vacuum energy [10].  

Having formulated the conditions for the eigenfrequencies of the
system it now remains to be shown how Cauchy's theorem can be applied to the
evaluation of the Casimir energy.  In analogy to ref. 7, the eigenfrequency
equations are defined as

\begin{eqnarray*}
f_l^{(1)}(z) & = & j_l(z) \\
f_2^{(2)}(z) & = & j_l(z)+ \tan{\delta_l(z)}n_l(z) \\
f_l^{(3)}(z) & = & {d\over dz}[zj_l(z)] \\
f_l^{(4)}(z) & = & {d\over dz}\biggl[z[j_l(z)-cot\delta_l(z)n_l(z)]\biggr]
\end{eqnarray*}
where $z=\omega a$ and 
$$\delta_l(z)=z(R/a)-{l\pi \over 2}.$$
Another useful definition is what might be termed the f-product, or 
$$f_l(z)=z^2\prod_{\alpha}f_l^{(\alpha)},$$
where a factor of $z^2$  has been included for convenience in order to make
$f_l(z)$ finite at $z=0$.  Defining $z$ to be a complex variable,
$f_l(z)$ is seen to be an analytic function of $z$, the zeros of which 
correspond to the eigenfrequencies of the system [11].

It follows from Cauchy's theorem that for two functions $f_l(z)$ and $\phi (z)$
analytic within a closed contour C in which $f_l(z)$ has isolated zeros at
$x_1,x_2,...x_n$,
$${1\over 2\pi i}\oint_C \phi (z){f_l^{\prime}(z)\over f_l(z)}dz=\sum_i
\phi (x_i).$$  
Choosing $\phi (z)=z e^{-\sigma z}$ where $\sigma $ is a real positive
constant thus leads to
 
\begin{equation}
{1\over 2\pi i}\oint_C e^{-\sigma z}zdz{d\over dz}\ln f_l(z)=\sum_i
z_i e^{-\sigma z_i}.
\end{equation}
Upon combining Eq.(8) with Eq.(2) the Casimir energy becomes

\begin{equation}
E_c=\lim_{\sigma\to 0}{1\over 2\pi ia}\sum_{l=1}^{\infty}
(l+{1\over 2})\oint_C dz e^{-\sigma z}z{d\over dz}\ln f_l(z).
\end{equation}
It should be noted that the factor of $e^{-\sigma z}$ plays the role of a
cutoff function which effectively suppresses the high frequency
contributions to the Casimir energy.  While such a cutoff function is generally 
invoked in calculations based on parallel plate geometry, it was not
employed in refs. 6 and 7.   It will be seen, however, that it plays an
important role in ensuring that all integrals encountered in Eq. (9) are well
defined.   

Before the calculation of the Casimir energy can be completed, it is necessary
to specify a contour for the integration.  An appropriate contour C 
(displayed in Fig.1) can be conveniently broken into three parts.  
These consist of a circular
segment $C_{\Lambda}$ and two straight line segments $\Gamma_1$ and $\Gamma_2$.
Since the $\Gamma$ contours
are oriented at a nonzero angle $\phi$ with respect to the
imaginary axis, it follows that the contribution to $C_{\Lambda}$ is, by virtue 
of the cutoff factor, bounded by $exp(-\sigma
\Lambda sin\phi)$ where $\Lambda$ is the radius of the circular arc.  

\begin{figure}[t]
\begin{center}
\mbox{\epsfig{figure=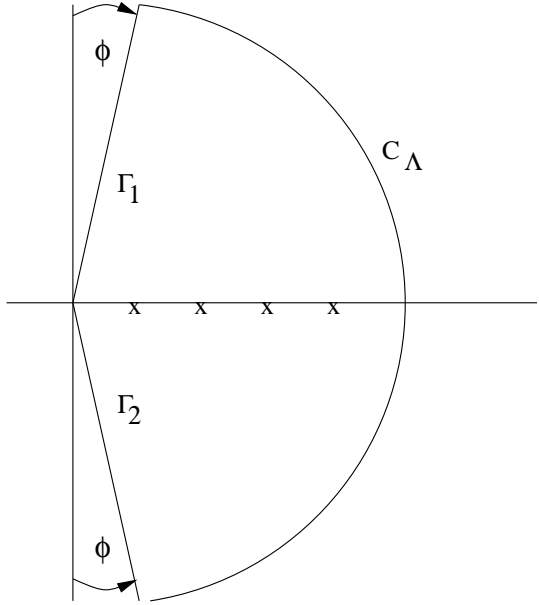,width=8.0cm,angle=270}}
\vspace*{\baselineskip}
\caption[]{Contour integral in complex plane.
}
\label{f1}
\end{center}
\end{figure}
\noindent Since the
logarithm in the Casimir energy expression grows at most algebraically, it
follows that the contribution to $C_{\Lambda}$ vanishes exponentially 
in the limit of large $\Lambda$ provided that $\phi \neq 0$.

Along $\Gamma_1$ and $\Gamma_2$
the quantity $tan\delta_l$ becomes $i$ and $-i$
respectively for sufficiently large $R$.  Thus on $\Gamma_1$, where $y=|z|$,

\begin{eqnarray*}
f_l^{(2)}(z) & \to & h_l^{(1)}(iye^{-i\phi}) \\
f_l^{(4)}(z) & \to & {d\over dy}[yh_l^{(1)}(iye^{-i\phi})],
\end{eqnarray*}
with the corresponding terms for $\Gamma_2$ obtained by complex conjugation.  
This
allows $f_l(z)$ to be written, up to an overall normalization, in terms of the 
modified Bessel functions as 
                      
\begin{eqnarray*}
f_l(z)\biggr|_{\Gamma_1} & = & I_\nu(ye^{-i\phi})K_\nu(ye^{-i\phi}) \nonumber \\
& & [{1\over 2}I_\nu(ye^{-i\phi})+ye^{-i\phi}I_\nu^{\prime}(ye^{-i\phi})] 
\nonumber \\
& & [{1\over 2}K_\nu(ye^{-i\phi})+ye^{-i\phi}K_\nu^{\prime}(ye^{-i\phi})]
\end{eqnarray*}          
where $\nu =l+{1\over 2}.$
Upon noting that the contributions along $\Gamma_1$ and $\Gamma_2$ are complex
conjugates of each other,
the Casimir energy becomes
$$E_c=-{1\over\pi a}\sum_{l=1}^{\infty} (l+{1\over2}) {\bf Re} 
e^{-i\phi}\int^{\infty}_0 dy
exp(-i\sigma y e^{-i\phi}) y {d\over dy}ln f_l(ye^{-i\phi}) .$$
Straightforward manipulation [5] of the argument of the logarithm and rescaling
of the integration variable allows this to be recast in the form

\begin{equation}
E_c=-{1\over \pi a}\sum_{l=1}^{\infty} 
(l+{1\over 2})^2 {\bf Re} e^{-i\phi}\int^{\infty}_0
dy exp(-i\nu\sigma ye^{-i\phi}) y {d\over
dy}ln\biggl[1-[\lambda_l(y\nu e^{-i\phi})]^2\biggr]
\end{equation}
where 

$$\lambda_l(y)={d\over dy}[yI_\nu(y)K_\nu(y)].$$
Upon using the uniform expansions of $I_{\nu}(\nu y)$ and $K_{\nu}(\nu y)$ for
large $\nu$ [12] it is found that the integral over $y$ is finite for
$\sigma =0$.  Adding and subtracting the asymptotic form of the 
integrand for large $\nu$, the Casimir energy assumes the form
$$E_c=E_{fin} +E_{\sigma},$$
where the finite part is 
 
$$E_{fin}={1\over \pi a}\sum_{l=1}^{\infty}\int_0^{\infty}dy\biggl[ (l+{1\over
2})^2ln[1-(\lambda (\nu y))^2]+{1\over 4}(1+y^2)^{-3} \biggr] $$
and the cutoff dependent part is
 
$$E_{\sigma}={1\over 4\pi a}\sum_{l=1}^{\infty} {\bf Re} 
e^{-i\phi}\int_0^{\infty}dy
exp(-i\nu\sigma ye^{-i\phi})y{d\over dy}(1+y^2e^{-2i\phi})^{-3}.$$
In arriving at the $\phi$ independent form for $E_{fin}$ one has in succession
taken the limit $\sigma\to 0$ and replaced $y$ with $ye^{i\phi}$.  The resulting
integral over the interval $0<y<e^{-i\phi}\infty$ is then trivially seen to
be equivalent to the corresponding integral over the real interval
$0<y<\infty$.  The quantity $E_{fin}$ has been evaluated numerically in ref. [6]
and consequently no attempt is made here to obtain that result independently.  

Although the integral which appears in $E_{\sigma}$ is finite even for
$\sigma =0$, the limit cannot be taken at this point since it would yield the
divergent result
$$E_{\sigma}\bigg|_{\sigma=0}=-{3\over 64}\sum_{l=1}^{\infty}(l+{1\over 2})^0.$$
Consequently, $E_{\sigma}$ is to be evaluated for finite $\sigma$ before
taking the limit of vanishing cutoff.  Since this is clearly a delicate
limiting process, some care is warranted.  It should first of all be noted 
that the cutoff factor used here was initially introduced as a real exponential
prior to its emergence in $E_{\sigma}$ as the term 
$exp(-i\nu\sigma ye^{-i\phi})$.  Thus there is a damping term of 
the form $exp(-\nu\sigma ysin\phi)$.  
This may be contrasted with ref. [6] which objects to using 
an explicit cutoff but finds, by using a temporal point separation, that an
oscillatory factor of the form $e^{i\epsilon\nu y}$ can appear in 
$E_{\sigma}$.  It is significant
that such a term (unlike that found here) has no residual exponential damping
effect. This has led to a criticism of the ref. 6 result by Candelas [8] who
asserts that because the waves of different $l$ all contribute an amount of 
equal absolute value to the Casimir energy, an explicit cutoff must be
introduced in the sum over $l$.  His result is that such an insertion 
adds a divergent term to the Casimir energy that is 
proportional to $\int_0^{\infty}d\nu$ (where $\nu$ is a
frequency).  Since, however, the calculation presented
here has included a cutoff {\it ab initio} there should be no need for the 
inclusion
of an additional cutoff of the type used in [8].  This can now be verified 
by explicitly carrying out the summation over $l$ to give

$$E_{\sigma}=-{3\over 2\pi a} {\bf Re} 
e^{-i\phi}\int_0^{\infty}dy{y^2exp(-i\sigma
ye^{-i\phi}/2)\over (1+y^2e^{-2i\phi})^4}{e^{-2i\phi}\over exp(i\sigma
ye^{-i\phi})-1}.$$
For nonzero $\phi$ the sum is well defined and thus requires no
additional cutoff factor.  Upon expanding the $\sigma$ dependent terms and
dropping those which go to zero as some positive power of $\sigma$ the above
expression reduces to 
$$E_{\sigma}={3\over 2\pi
a}{\bf Re}\int_0^{\infty}dy{y^2e^{-3i\phi}\over(1+y^2e^{-2i\phi})^4}[1+{i\over\sigma
ye^{-i\phi}}].$$
As before, let $y\to e^{i\phi}y$ and perform a rotation of the resulting
contour.  This allows the term singular in $\sigma$ to be dropped (it is purely
imaginary) and leaves the finite result
$$E_{\sigma}\bigg|_{\sigma =0}={3\over 64a}.$$
Upon combining this with the numerical result of ref. [6] for $E_{fin}$,
the expression for the Casimir energy is obtained, namely
$$E_c=0.09235/2a.$$

As a final comment concerning the conducting sphere, it should be noted
that the method presented here can equally well be applied to the Dirichlet and
Neumann spheres.  The results for these cases can be rigorously obtained 
using the technique of mode summation in
conjunction with Cauchy's theorem.  As it would consist merely of a 
rewriting of ref. [7] using the approach developed here, this is left as
an exercise for the interested reader.   

\medskip

\begin{center}
{\bf III. Casimir Effect of the Electromagnetic Circle}\\
\end{center}

A natural question which arises subsequent to a consideration of the
electromagnetic Casimir effect for a sphere is whether similar
results can be obtained in higher and lower dimensions.  As will be shown here
this question can readily be answered at least in the latter case.  Although
the
electromagnetic field is trivial (i.e., there is no photon) in one spatial
dimension, for the two dimensional case a procedure analogous to that of the
preceding section can be carried out.
This requires the extraction of appropriate boundary
conditions from the field equations.  The latter are contained in the covariant
equations

\begin{equation}
F^{\mu \nu}=\partial^{\mu}A^{\nu}-\partial^{\nu}A^{\mu}
\end{equation}

\begin{equation}
\partial _{\nu}F^{\mu \nu}=J^{\mu}
\end{equation}
where $J^{\mu}=(J_i, \rho)$ with $i=1,2$.  From (11) it follows that
$$B=\mbox{\boldmath $\nabla$}\times {\bf A}$$
and
$${\bf E}=-\partial_0{\bf A}-\mbox{\boldmath $\nabla$} A^0$$
where $B$ is the (scalar) magnetic field and ${\bf E}$ is the (two-vector)
electric field. Correspondingly, Eq.(12) yields
$$\mbox{\boldmath $\nabla$}\cdot {\bf E}=\rho$$
and
$$-\partial_0{\bf E}+\mbox {\boldmath $\overline \nabla$}B={\bf J}$$
where $(\overline \nabla)_i=\epsilon_{ij}\nabla_j$.

In the case that all fields have time dependence $e^{-i\omega t}$ these
equations imply that in current free regions
$$(\mbox{\boldmath $\nabla$}^2 +\omega^2)B=0.$$
Thus the $B$ field can be written in the form
$$B(r,\phi)=\sum_{m=-\infty}^{\infty}[a_m J_m (kr) + b_m N_m(kr)]e^{im\phi}$$
with the corresponding ${\bf E}$ field given by
$$-i \omega {\bf E}=\mbox{\boldmath $\overline \nabla$}B.$$
Since
$$\mbox{\boldmath $\nabla$}\times{\bf E}=-\partial_0B,$$
it follows that for a conducting circle of radius $a$ the appropriate boundary
condition is ${\bf r}\times{\bf E}|_{r=a}=0$, or equivalently,
${\partial\over \partial r}B|_{r=a}=0$.  In other words, the Casimir effect
in (2+1) dimensional QED is equivalent to that of a scalar field satisfying
Neumann boundary conditions.

For the inside modes of the system it follows that the eigenfrequencies
are determined by the condition
$${d\over dr}J_m(\omega r)\bigg|_{r=a}=0.$$
The outside modes require that in analogy to the three dimensional case the
system be enclosed in a large circle of radius $R$ which is subsequently taken
to infinity.  Thus the boundary condition in the exterior region takes the form
$${d\over dr}[J_m(\omega r)+tan\delta_lN_m(\omega r)]\bigg|_{r=a}=0$$
where $\delta_l$ is given by Eq.(7).  Upon repeating the procedure of the
preceding section it is found that the Casimir energy of the circle has the
form
$$E_c=-{1\over 2\pi a}\sum_{m=-\infty}^{\infty}{\bf Re} e^{-i\phi}\int_0^{\infty}
dyexp(-i\sigma ye^{-i\phi})y{d\over dy}ln
[yI_m^{\prime}(ye^{-i\phi})K_m^{\prime}(ye^{-i\phi})].$$
The question which needs to be addressed is whether upon performing the sum
over $m$ a finite result for the Casimir energy can be obtained in the $\sigma
\to0$ limit.  Using the uniform expansions for the modified Bessel
functions for large $|m|$ and considering only those $m$ values with absolute
value greater than some large integer $M$, it is found that
$$E_c\to {1\over 2a}\sum_{m=M}^{\infty}me^{-m\sigma}.$$
This yields the divergent result
$$E_c\to {1\over 2a}{1\over \sigma^2}$$
in the limit of vanishing $\sigma$, a result at variance with the claim in ref.
[13] of a finite Casimir energy.  On the other hand, it is not unexpected in
view of the divergence found by Bender and Milton [14] for the two
dimensional Dirichlet Casimir effect.

\newpage

\begin{center}
{\bf IV. Conclusion}\\
\end{center}

Since the inception of this subfield, there have been a number of approaches 
developed for calculating
the Casimir energy of a conducting sphere in vacuum. 
The first successful evaluation consisted of analytically summing 
the individual mode contributions.  This method, closely modelled after the
much simpler parallel plate problem, was the most obvious at the time but 
included tedious algebraic manipulations.  More advanced methods followed which
rendered the involved mode summation approach obsolete  prior to its
re-examination in ref. 7.  The introduction of Cauchy's theorem 
(replacing the mode summation with an integration in the complex frequency 
plane) opened the door to a revitalization of this method. 
However, the formalism was based upon an improper treatment of the external
modes which led to their being complex and finite in number rather than real
and infinite as required.  A correct set of
equations for the exterior modes was obtained upon placing the entire system in
a second conducting sphere whose radius $R$ was eventually allowed to go to
infinity.

A second problem associated with the Cauchy integral method consisted in the
fact that the neglect of the semicircular part of the contour integral was
problematical.  This  difficulty was eliminated by invoking a cutoff function 
(as in the original parallel plate calculation), and by introducing
simultaneously a kink into the integration contour.  This allowed the integral
to be evaluated along the imaginary axis.  It
also led to a mathematically well defined summation over partial waves which
eliminated the need for a cutoff of the type proposed by Candelas.  Thus an
unobservable but divergent contribution to the Casimir energy which he claimed
to be essential was found not to be relevant.  It was noted that the method
developed here applies equally well to the Dirichlet and Neumann spheres, 
although no explicit calculations have been carried out.  Finally, an extension 
to the case of (2+1) dimensional QED was noted with divergent results being 
obtained in the limit of vanishing cutoff.

\medskip

\begin{center}
{\bf Acknowledgments}\\
\end{center}
\noindent This work was supported in part by the U.S. Department of 
Energy Grant No.DE-FG02-91ER40685.                     

\medskip
\noindent $^*$ Electronic address: hagen@urhep.pas.rochester.edu


%
%

\newpage
\noindent References

\begin{enumerate}                                 

\item H. B. G. Casimir, Proc. Koninkl. Ned. Akad. Wetenschap. {\bf 51}, 793
(1948).
\item H. G. B. Casimir, Physics {\bf 19}, 846 (1956).
\item T. H. Boyer, Phys. Rev. {\bf 174}, 1764 (1968).
\item B. Davies, J. Math. Phys. {\bf 13}, 1324 (1972).
\item R. Balian and B. Duplantier, Ann. Phys. (N.Y.) {\bf 112}, 165 (1978).
\item K. A. Milton, L. L. DeRaad, Jr., and J. Schwinger, Ann. Phys.
(N.Y.) {\bf 115}, 388 (1978).
\item V. V. Nesterenko and I. G. Pirozhenko, Physical Review D {\bf 57}, 1284
(1997).
\item P. Candelas, Ann. Phys. (N.Y.) {\bf 143}, 241 (1982).
\item J. A. Stratton, {\it Electromagnetic Theory} (McGraw-Hill, New York,
1941).
\item It is also of interest to note that the boundary conditions of ref. 7
imply [9] that there are only $2l+1$ exterior modes for each $l$ even though
Eqs. (3) and (5) allow an infinite number of interior modes.
The approach presented here, however, preserves the symmetry between the
interior and exterior regions (i.e., there are an infinite number of real 
eigenfrequencies in each case).  
\item The fact that there are no other zeros of $f_l(z)$ for $Im z\neq 0$ is
perhaps most easily seen from the fact that the eigenfrequencies of this system
correspond to those of a free quantum mechanical particle in three dimensions
which is confined to the interior of a sphere of radius $a$ or to the region
between two concentric spheres of radii $a$ and $R$, subject to either
Dirichlet or Neumann boundary conditions.  Since all the energy eigenvalues of
such a system are known to be real, it follows that $f_l(z)$ has zeros only on
the real axis.
\item {\it Handbook of Mathematical Functions}, edited by M. Abramowitz and I.
A. Stegun (National Bureau of Standards, Washington, D. C., 1964), Chap. 9.
\item K. A. Milton and Y. J. Ng, Phys. Rev. D {\bf 46}, 842 (1992).
\item C. M. Bender and K. A. Milton, Phys. Rev D{\bf 50}, 6547 (1994).

\end{enumerate}

\end{document}